\documentclass[10pt]{iopart}
\usepackage{dcolumn}
\usepackage{bm}
\usepackage{graphicx}
\expandafter\let\csname equation*\endcsname\relax
\expandafter\let\csname endequation*\endcsname\relax
\usepackage{amsmath}
\usepackage{latexsym}
\usepackage{amsfonts}
\usepackage{amssymb}
\usepackage{array}
\usepackage{epsfig}
\usepackage{txfonts}
\usepackage{color}
\usepackage[colorlinks=true,linkcolor=blue,urlcolor=blue,citecolor=blue,pdfusetitle]{hyperref}
\usepackage{bm}
\usepackage{setspace}
\usepackage{braket}
\usepackage[utf8]{inputenc}
\usepackage[percent]{overpic}
\usepackage[dvipsnames]{xcolor}
\begin{document}

\title{A many-body heat engine at criticality}

\author{Thom\'{a}s Fogarty and Thomas Busch }
\address{Quantum Systems Unit, Okinawa Institute of Science and Technology Graduate University, Onna, Okinawa 904-0495, Japan}
\ead{thomas.fogarty@oist.jp}

%\author{Thomas Busch}
%\affiliation{Quantum Systems Unit, Okinawa Institute of Science and Technology Graduate University, Onna, Okinawa 904-0495, Japan}

\date{\today }

\begin{abstract}
    We show that a quantum Otto cycle in which the medium, an interacting ultracold gas, is driven between a superfluid and an insulating phase can outperform similar single particle cycles. The presence of an energy gap between the two phases can be used to improve performance, while the interplay between lattice forces and the particle distribution can lead to a many-body cooperative effect. Since finite time driving of this cycle can create unwanted non-equilibrium dynamics which can significantly impair the performance of the engine cycle, we also design an approximate shortcut to adiabaticity for the many-body state that can be used to achieve an efficient Otto cycle around a critical point.
\end{abstract}

\maketitle
\ioptwocol
\section{Introduction}
The almost unmatched precision of controlling and measuring cold atomic systems provided by recent experiments has made them forerunners in the area of quantum simulations \cite{OpticalLatticeLewenstein:07,bloch_quantum_2012,Gross995}. In particular their many-body aspect and the ability to create out-of-equilibrium situations in a controlled way has led to paradigmatic results that are beyond even advanced numerical simulations \cite{RevModPhys.83.863}.  They therefore  offer an exciting testbed for exploring ideas in quantum thermodynamics \cite{SebStevQTBook}, ranging from  insights into the understanding of work and heat at the quantum level to the operation of quantum heat engines (QHE) and refrigerators \cite{ottoX,Hallwood,LutzPRLHeatEngine,LutzScience,Kosloff2017,ChenCampo2019}. Describing such machines taking fundamental quantum effects into account has already led to a number of unexpected results and can allow one to achieve certain advantages over comparable classical systems. In recent years this has been shown for machines operating across quantum phase transitions \cite{CampisiNC,FuscoPRE,QPTHE1,QPTHE2,QPTHE3,Revathy2020,Abiuso2020}, using squeezed baths as quantum environments \cite{squeeze1,squeeze2,squeeze3,squeeze4,GershonNC}, or exploiting the cooperative effects of many-body quantum systems \cite{Uzdin2016,Campaioli2017,Vroylandt_2017,adolfo3,delCampoEntropy,DeffnerEnt,Niedenzu_2018,Ferraro2018}. 

However, the description of interacting many-particle systems at finite temperatures is a non-trivial problem and solvable models only exist in restricted circumstances that are often not experimentally realistic. One notable exception to this are the recently realised Tonks-Girardeau (TG) gases of strongly interacting bosons in effectively one-dimensional settings \cite{paredes_tonksgirardeau_2004,Kinoshita1125}, where exact solutions can be found using the Bose-Fermi mapping theorem at finite temperatures \cite{girardeau1960relationship,Lenard1966,Vignolo2013,Atas2017}. Therefore they lend themselves to exact studies of thermodynamical machines.

In this work we consider a Tonks-Girardeau gas in a box and realise the compression and expansion strokes a heat engine requires by the switching on and off of an optical lattice potential. This changes the one-dimensional volume the system has available and also leads to significant changes in the energy spectrum. Even more, in such a system the particle filling statistics plays an important role, as at low temperature and unit filling an insulating phase forms as soon as an infinitesimally weak lattice potential is applied \cite{buchler2003commensurate,haller2010pinning}. This phase transition is called the pinning-transition and it is signalled by the appearance of an energy gap in the spectrum.  One can therefore drive a quantum Otto cycle between the superfluid and insulating phases by simply switching the lattice on and off. As the operation of the engine cycle is dependent on the energy spectrum of the particles, the presence of the energy gap at the quantum critical point can drastically change the engine performance. Furthermore, due to the competing influence of the lattice potential and the particle interactions, nontrivial energy spectra can be achieved that may exhibit a many-body cooperative effect on the engine cycle. This can be quantified by comparing the many-body quantum heat engine (mQHE) with an equivalently sized ensemble of non-interacting single particle quantum heat engines (sQHE) \cite{adolfo3}. 

Of course, any realistic implementation of a QHE cycle must be carried out on a finite timescale, which can have a negative impact on the resulting engine performance. If the cycle is performed too quickly, the excitation of non-equilibrium states may act as a form of inner friction due to the irreversible nature of the dynamics, thereby reducing performance \cite{MossyNJP,FuscoPRX}. While adiabatic dynamics preserve the reversibility of the cycle through the slow driving of the quantum state, the long timescales required result in negligible output power. To achieve both, engine cycles that are efficient and fast, one can employ the techniques of shortcuts to adiabaticity (STA), which allow for adiabatic dynamics on finite timescales \cite{GooldSciRep,Abah2018,PRXAdolf,LutzPRLHeatEngine,CampbellPRA,AbahPRE:19,MGaraot,polkovnikov,delCampoEntropy, Abah19}. However, since the driven dynamics of our interacting many-particle system encompasses the quantum critical point at the pinning transition, standard STA approaches cannot be easily employed. We therefore derive and implement a many-body STA using a variational approach \cite{LewisSTA,Xu:20}, which, although approximate in nature, improves the performance of the engine when compared to a non-optimised cycle. 

\section{Methods}
\begin{figure}[tb]
    \includegraphics[width=\columnwidth]{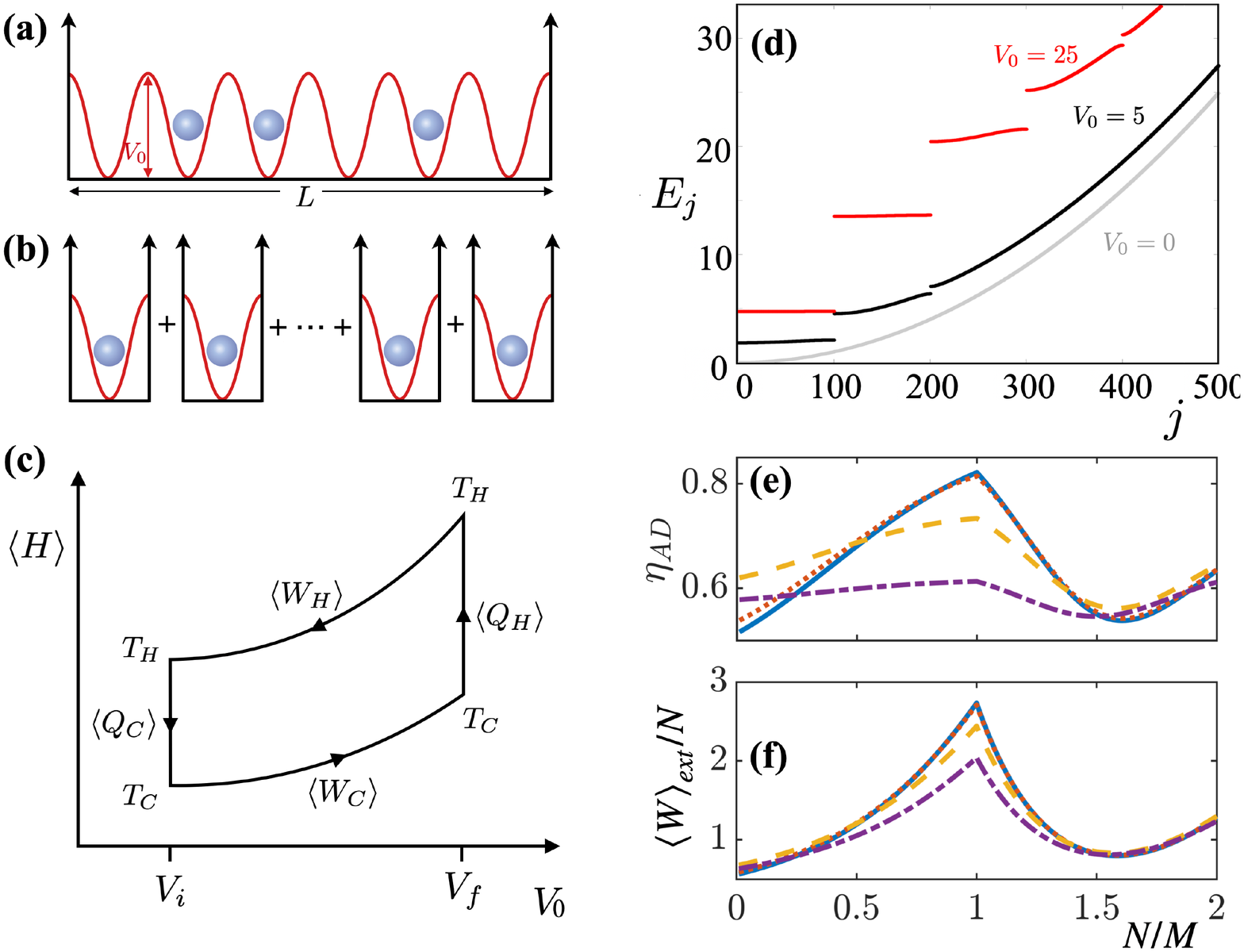}
    \caption{(a) Schematic of the setup for the mQHE and (b) the sQHE. (c) Schematic of the Otto cycle. (d) Single particle spectrum $E_j$ of a lattice with $M=100$ wells and $V_0=0$ (light grey), $V_0=5$ (black) and $V_0=25$ (red).  (e) Efficiency and (f) work output for an adiabatic cycle as a function of the filling ratio $N/M$, with $T_C=0$, $T_H=5$ and $V_f=50$. Different values of $V_i$ are shown, $V_i=0$ (blue solid), $V_i=1$ (red dotted), $V_i=5$ (yellow dashed) and $V_i=10$ (purple dot-dashed). The work output is scaled with the number of particles $N$.}
    \label{Fig:Schematic}
\end{figure}

The system we consider consists of a gas of $N$ particles of mass $m$ which are trapped in an effectively one dimensional box potential, $V_\text{B}(x)$, of length $L$ with infinitely high walls. The single particle Hamiltonian is given by 
\begin{equation}
    H=-\frac{\hbar^2}{2m}\nabla^2+V_\text{B}(x) + V_l(x,t)\;,
    \label{eq:SingleParticleHamiltonian}
\end{equation}
where we have also included a time-dependent optical lattice potential of the form $V_l(x,t)=V_0(t) \cos^2(k_0 x+\phi)$ (see Fig.~\ref{Fig:Schematic}(a)). The lattice vector is given by $k_0=M\pi/L$ and $M$ is the number of wells. We choose $\phi=0$ for $M$ even and $\phi=\pi/2$ for $M$ odd to ensure that there are no half lattice sites at the edge of the box. We also choose to fix $k_0$ and scale the size of the box potential to change the number of lattice sites.

The eigenstates, $\psi_n(x)$ (which we calculate through exact diagonalization), of the Hamiltonian \eqref{eq:SingleParticleHamiltonian} can be used to describe a gas of spinless fermions via the Slater determinant $\Psi_{F}(x_1,x_2,\dots,x_{N})=\frac{1}{\sqrt{N!}}\det^{N}_{n,j=1}[\psi_{n}(x_{j})]$, which can be mapped onto a TG gas of hard-core bosons after appropriate symmetrization as $\Psi_{B}(x_1,x_2,\dots,x_{N})=\prod_{1\le i < j \le N}\mbox{sgn}(x_{i}-x_{j})\Psi_{F}(x_{1},x_{2},\dots,x_{N})$ \cite{girardeau1960relationship,Girardeau2000}. This duality can be understood by realising that the spatial distribution of the fermions is governed by a pseudo-interaction implied by the Pauli exclusion principle (Fermi pressure), which has the same effect as the strongly repulsive interaction present in the TG gas. Their respective densities are therefore trivially identical, and this equivalence also extends to their thermodynamic behaviours which are governed by the Fermi-Dirac occupation factors, $f_n=[e^{(E_n-\mu)/k_BT}+1]^{-1}$ (with $E_n$ the eigenenergies, $\mu$ the chemical potential and $k_B$ the Boltzmann constant) \cite{Lenard1966,Atas2017b,Atas2017}. This implies that the engine cycles will be identical as well. In the following we will scale all energies in units of the lattice recoil energy, $E_R=\hbar^2k_0^2/(2m)$, and temperature in units of $E_R/k_B$.

We consider a quantum Otto cycle (see Fig.~\ref{Fig:Schematic}(c)) driven between two lattice depths, $V_0=\lbrace V_i, V_f \rbrace$, at different lattice filling ratios, $N/M$. The cycle consists of four strokes: (i) isentropic \textit{compression} (lattice raising from depth $V_i$ to $V_f$) over a time $t_1$ at fixed temperature $T_C$; (ii) weak coupling to a thermal bath at temperature $T_H>T_C$ during a time $t_2$; (iii)  isentropic \textit{expansion} (lattice lowering from depth $V_f$ to $V_i$) over a time $t_3$; and (iv) weak coupling to a thermal bath at temperature $T_C$ for a time $t_4$. During the isentropic compression and expansion strokes we assume the system is isolated from the respective thermal reservoirs. 

We consider a reversible cycle where the dynamics of the quantum state are sufficiently slow so as to be considered adiabatic (denoted by the subscript $AD$). The work done during the isentropic strokes can be calculated from the difference in energy between the many-body states at lattice depths $V_i$ and $V_f$ at the different temperatures, $ \langle W_C\rangle =\langle H_{T_C}(V_f) \rangle - \langle H_{T_C}(V_i) \rangle$ and $\langle W_H\rangle=\langle H_{T_H}(V_i) \rangle - \langle H_{T_H}(V_f) \rangle$, with $\langle H \rangle = \text{Tr}(H \rho)$ being the expectation value of the energy of the thermal states. The heat exchanged with the cold and hot baths is the given by $ \langle Q_C \rangle=\langle H_{T_C}(V_i) \rangle - \langle H_{T_H}(V_i) \rangle$ and $\langle Q_H \rangle=\langle H_{T_H}(V_f) \rangle - \langle H_{T_C}(V_f) \rangle$ and the efficiency and output power can be calculated as
\begin{equation}
  \eta_{AD}=-\frac{\langle W_C\rangle+\langle W_H\rangle}{\langle Q_H \rangle},\quad
  P_{AD}=-\frac{\langle W_C\rangle+\langle W_H\rangle}{\tau}\;,
  \label{eq:effpow}
\end{equation}
where $\langle W \rangle_{\text{ext}}=-\left( \langle W_C\rangle+\langle W_H\rangle\right)$ is the work output and $\tau=t_1+t_2+t_3+t_4$ is the duration of the cycle. 

\section{Results}
\subsection{Adiabatic cycle}
For the performance of the engine the filling fraction $N/M$ plays an important role. At $T_C=0$ and for an incommensurate filling, $N\neq M$, the particles are delocalized in the lattice and can move within the box. However, for a commensurate filling, $N=M$, a pinning transition occurs for any infinitesimal lattice strength, whereby each particle becomes more strongly localized at an individual lattice site, which significantly restricts its motion \cite{lelas2012pinning,Mikkelsen_2018}. The behaviour of this insulating phase is then determined by the energy gap in the single particle spectrum (see Fig.~\ref{Fig:Schematic}(d)) which has a size of approximately $V_0/2$ for shallow lattices and $2\sqrt{V_0}$ for deep lattices \cite{buchler2003commensurate}. The differences in the accessible single particle excitation spectrum for $N/M$ therefore lead to different behaviours when running the engine and in Fig.~\ref{Fig:Schematic}(e,f) one can clearly see that peak performance is achieved at unit filling. At this point the particles in the cold adiabat fill the lowest energy band and as $V_0$ is increased the energy gap is widened. Thermal excitations induced by the hot bath then allow particles to jump the gap and therefore more energy can be extracted as the lattice depth is decreased along the hot adiabat. As the high performance regime is the interesting one for heat engines, we will focus on the case of unit filling in the following. 

\begin{figure*}[tbp]
    \includegraphics[width=2\columnwidth]{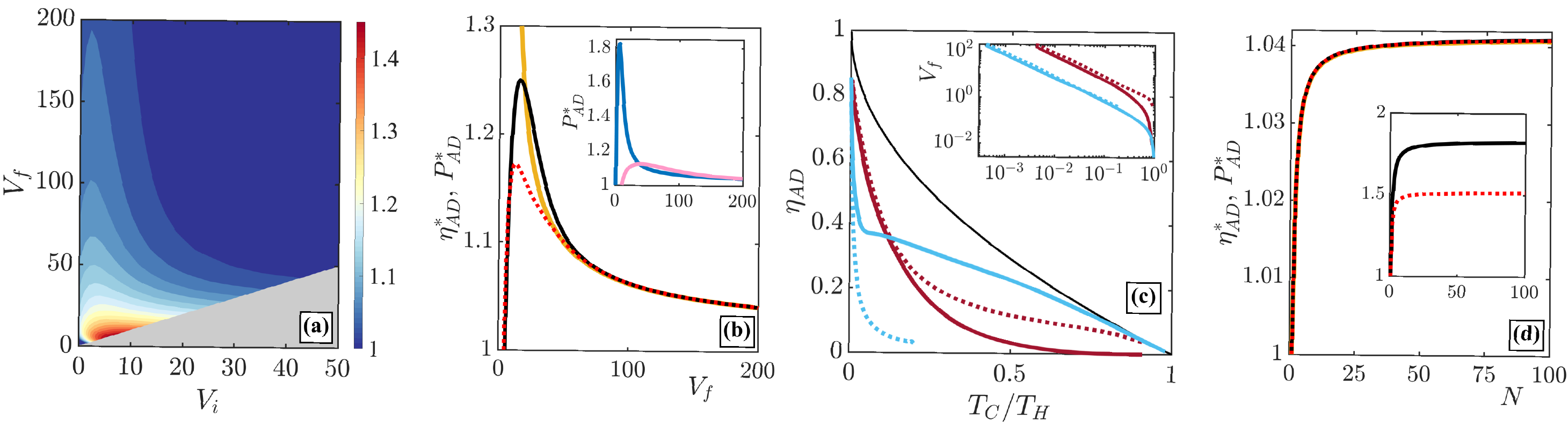}
    \caption{(a) Efficiency ratio $\eta^*_{AD}$ as a function of the initial lattice depth $V_i$ and the final lattice depth $V_f$. The system is at unit filling with $N=M=100$, while the temperatures of the cold and hot reservoirs are $T_C=0$ and $T_H=5$, respectively. The grey region indicates the parameter space where $V_i>V_f$, in which no work can be extracted from the cycle. (b) Numerical values for $\eta^*_{AD}$ (red dotted line) and $P^*_{AD}$ (black solid line) compared to the approximation given in Eq.~\eqref{eq:supp} (yellow solid line) as function of $V_f$ for $T_H=5$ with $N=M=100$. The inset shows $P^*_{AD}$ for $T_H=1$ (blue line) and $T_H=20$ (pink line). (c) Efficiency at maximum power (see text) and shown as a function of $T_C/T_H$ for two temperatures of the cold adiabat:  $T_C=0.01$ (light blue lines) and $T_C=0.1$ (dark red lines). The solid lines are for the mQHE, the dotted ones for the sQHE and the black line indicates the Curzon-Ahlborn limit. The inset shows the values of $V_f$ which correspond to the maximum power for each respective cycle (in log-log scale as a function of $T_C/T_H$). (d) Dependence of $\eta^*_{AD}$ (red dotted line) and $P^*_{AD}$ (black solid line) as a function of $N$ at unit filling with $T_H=5$ and with $V_f=200$. The numerical results are indistinguishable from Eq.~\eqref{eq:supp}. The inset shows $\eta^*_{AD}$ and $P^*_{AD}$ for $T_H=1$ and $V_f=5$.}
\label{fig:AD}
\end{figure*}

The advantage of exploiting the critical point in a many-body quantum heat engine (mQHE) with $N$ particles can be quantified by comparison with an ensemble of $N$ single-particle quantum heat engines (sQHE), see Fig.~\ref{Fig:Schematic}(b).  Each sQHE obeys the Hamiltonian given by Eq.~\eqref{eq:SingleParticleHamiltonian} with a box length of $L=\pi/k_0$, so that exactly one lattice well is present, $V_l=V_0 \cos(k_0 x + \pi/2)$. The Otto cycle is then carried out using the same lattice height and bath temperatures, however in the mQHE the final state is strongly influenced by the presence of the interparticle interactions and the periodicity of the optical lattice. %, while the energy of the state in the SPHE is only affected by the external trapping potentials. 
To quantify the difference between the mQHE and the sQHE, we calculate the ratio of their respective efficiencies and powers
\begin{equation}
 \eta^*(N)=\frac{\eta(N)}{\eta(1)},\quad
  P^*(N)=\frac{P(N)}{N P(1)}\, ,
\end{equation}
such that $\eta^*(N)>1$ and $P^*(N)>1$ indicate that the many-body state gives a performance boost \cite{adolfo3}.

In Fig.~\ref{fig:AD}(a) we show the efficiency ratio for an adiabatic cycle as a function of the lattice depths $V_i$ and $V_f$. For $V_f>V_i$ the cycle produces positive work and therefore acts as an engine. One can see that large many-body cooperative effects can be achieved in the regime where both lattices are weak, and where therefore the particles in the mQHE are still partially overlapping. This results in a non-trivial, non-flat single-particle energy spectrum (see Fig.~\ref{Fig:Schematic}(d)) and therefore in enhanced efficiency and power output over the sQHE. When both lattices are deep, $V_0\gtrsim 30$, the particles are highly localized in individual lattice sites and the single particle energy spectrum becomes degenerate forming flat bands. In this limit all many-body cooperative effects are lost and the mQHE becomes equivalent to the sQHE. Since for weak initial lattice depths, $V_i\lesssim 10$, the mQHE shows enhanced performance for a range of values of $V_f$, we will focus on this region of the parameter space in what follows, specifically considering the limiting case of initially having free particles ($V_i\rightarrow 0$).

Indeed, for mQHE cycles which operate at low reservoir temperatures, $\theta=\frac{E_R}{k_B T_H}\sqrt{V_f} >1$, and which ramp to deep lattices, $V_f\gg1$, it is possible to find an approximate expression for the many-body performance boost,
\begin{equation}
    \eta^*_{AD}(N)=P^*_{AD}(N)\approx1+\frac{1-1/N}{\Delta-\frac{3}{2}\left[ \coth\left(\theta\right)+1\right]}\;,
    \label{eq:supp}
\end{equation}
where $\Delta=2\sqrt{V_f}-1$ is the energy gap (see Supplemental Material for details). From this one can immediately see that at unit filling the mQHE will always outperform the sQHE once $V_f>4$. Furthermore, increasing the number of particles and reaching the state of double filling, $N=2M$, where the two lowest states of each lattice site are occupied, does not lead to improved performance. In the limits $\theta,M\rightarrow\infty$ the efficiency ratio can be written as $\eta^*(2M)\equiv\frac{\eta(2M)}{\eta(1)}\rightarrow\frac{1-4(\Delta-1)^{-1}}{1-3\Delta^{-1}}$, showing that for $V_f>1$ the efficiency of the sQHE is always larger than the mQHE at double filling, which is due to the anharmonicity of the individual lattice sites leading to reduced gaps between higher lying energy states.  

In Fig.~\ref{fig:AD}(b) we show the numerically obtained values of the ratios $\eta^*_{AD}(N)$ and $P^*_{AD}(N)$ as a function of $V_f$ in comparison to the approximation in Eq.~\eqref{eq:supp}. One can see that the exact ratios peak at lower values of $V_f$, which is due to the fact that the particles in the many-body state are still partially delocalized and therefore many-body cooperativity is stronger. For deeper lattices, $V_f\gtrsim50$, both ratios head towards one, as stronger localisation makes the lattice sites become effectively independent. The decay of the many-body advantage is well described by the approximation in Eq.~\eqref{eq:supp} (solid yellow line in Fig.~\ref{fig:AD}(b)) and given by a $1/\Delta$ dependence. While the decay is universal, the position and height of the maximum are depending on the other parameters of the system, in particular $T_H$ (see inset of Fig.~\ref{fig:AD}(b)). In general, a significant many-body advantage exists by operating the mQHE in weak lattices and at low temperatures when the commensurate system remains close to the quantum critical point. At higher temperatures the existing thermal energies diminish the importance of energy gap and the quantum criticality is washed out. To demonstrate this we show in Fig.~\ref{fig:AD}(c) the efficiency at maximum power (optimised over the lattice depth $V_f$, see inset in Fig.~\ref{fig:AD}(c)) for two different temperatures of the cold adiabat: one deep in the quantum regime $T_C=0.01$, which ensures that the system is in its ground state and therefore close to the quantum critical point; the other at a slightly higher temperature $T_C=0.1$, where the effect of the quantum criticality is reduced. When $T_H$ is small, the mQHE with the lower $T_C$ can be seen to be more efficient and close to the Curzon-Ahlborn efficiency, $\eta_{CA}=1-\sqrt{T_C/T_H}$, which is a good indicator of the performance of the Otto cycle \cite{CA1,DeffnerEnt,AbahPRE:19}. Furthermore, it is worth noticing that at higher $T_C$ the mQHE is outperformed by the sQHE as the thermal energy leads to less localisation within the box potential and the energy gap is washed out. Therefore, this quantum critical mQHE only shows enhanced performance at low temperatures when the system is close to the critical point. 

In deep lattices the power and efficiency ratios are equivalent for any number of lattice sites at unit filling ($N=M$) and they are exactly described by Eq.~\eqref{eq:supp} (see Fig.~\ref{fig:AD}(d)). As the many-body advantage is proportional to $(1-1/N)$, one can see a rapid increase in both quantities for increasing particle number until $N\sim10$, after which it asymptotically approaches $1+\left(\Delta-3\right)^{-1}$ in the thermodynamic limit. In more shallow lattices the efficiency and power ratios asymptotically reach different, but overall larger values, while the dependence on $N$ remains consistent with the behaviour observed for deep lattices (see inset of Fig.~\ref{fig:AD}(d)). Indeed, one does not need to create large many-body states to see a marked improvement in engine performance, rather only a few dozen particles are sufficient for observing the effects of many-body cooperativity in this system. 

\subsection{Finite time cycle}
While all the results above are obtained for a reversible cycle with undergoes adiabatic dynamics, this results in negligible power output due to the long timescales for each cycle, therefore necessitating fast engine cycles for finite power-output. However, fast driving through a critical point will inevitably result in non-adiabatic dynamics and irreversible work being produced, with the latter being defined as the difference between the average work of the non-adiabatic ($NA$) and adiabatic driving, $\langle W_{irr} \rangle=\langle W \rangle_{NA} - \langle W \rangle_{AD}$. This ultimately leads to reduced performance of the QHE \cite{campbell2016,Hoang2016,Deffner2017}. To explore the effect of irreversible work on the engine performance we numerically calculate the unitary dynamics of the single particle states during the compression and expansion strokes, $\psi_n(x,t)=e^{-\frac{i}{\hbar} \int_{0}^{t} H(t') dt'} \psi_n(x,0)$, describing the insertion and removal of the optical lattice over a finite time $t_f$ (in units of $2\pi/E_R$). We parametrise the lattice strength as $V_{\lambda}(t)=\lambda(t)V_f\cos^2(k_0 x)$, with $\lambda(t)=t^3/t_f^3[1+3(1-t/t_f) + 6(1-t/t_f)^2]$, which %While this choice may not be optimal for this system (see \cite{Bloch2007,Masuda2014,Troyer2015,Zhou_2018}), it 
is sufficient to explore the dynamical properties of a finite time engine stroke, but is not necessarily optimal for the system \cite{Bloch2007,Masuda2014,Troyer2015,Zhou_2018}). As our focus is on the non-adiabatic dynamics initiated by the lattice ramp we will neglect the dynamics during the coupling to the different heat baths and assume that the thermalization times $t_2$ and $t_4$ are much shorter than the times for the work strokes $t_1$ and $t_3$ \cite{GooldSciRep,LutzArXiv}. Taking $t_1=t_3\equiv t_f$, the total time for the cycle is $\tau\approx 2 t_f$. 

\begin{figure}[tbp]
    \includegraphics[width=\columnwidth]{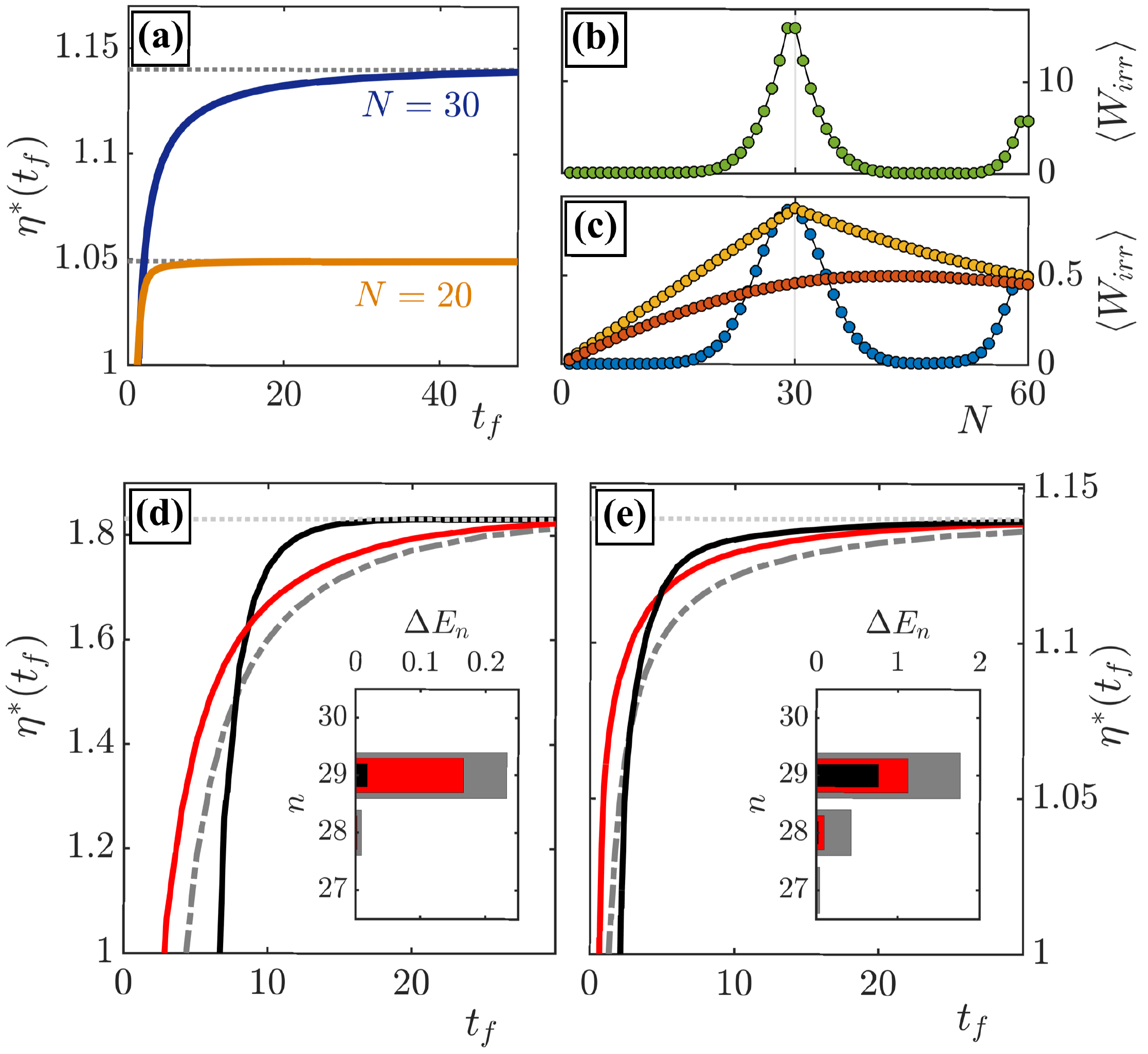}
    \caption{(a) Efficiency ratio $\eta^*$ as a function of ramp time $t_f$ for $M=30$ with $V_f=25$ and $T_H=5$ ($P^*$ behaves similarly and is not shown) at the pinning transition point ($N=30$) and away from it ($N=20$). (b-c) Irreversible work created during a process with a ramp time of $t_f=2$ as a function of total particle number $N$. (b) $\langle W_{irr} \rangle$ for ramping on the lattice at $T_C=0$, (c) $\langle W_{irr} \rangle$ for ramping off the lattice at $T_H=0$ (blue), $T_H=0.5$ (yellow) and $T_H=5$ (red). Efficiency ratio for (d) $V_f=5$ and $T_H=0.5$ and (e) $V_f=25$ and $T_H=5$ after implementing the non-optimised ramp $V_{\lambda}(t)$ (grey dot dashed), and the STAs $\langle V^\text{STA}(t) \rangle$ (red) and $V_{M-1}^\text{STA}(t)$ (black). Insets: energy difference between the non-adiabatic and adiabatic single particle energies after each ramp, $\Delta E_n=E_n^{NA}-E_n^{AD}$, taking $t_f=15$ and with the color code matching that of the larger panels.} 
\label{fig:ref_ramp}
\end{figure}

For a finite time cycle at commensurate filling the efficiency only slowly approaches the adiabatic efficiency (see Fig.~\ref{fig:ref_ramp}(a)) due to the large amount of irreversible work created when driving the system at the pinning transition (see Fig.~\ref{fig:ref_ramp}(b-c)). In comparison, incommensurate fillings produce significantly less irreversible work as excitations are far from the energy gap and therefore the adiabatic efficiency can be reached for significantly shorter ramp times. Also note that more irreversibility is created during the raising of the barrier compared to the lowering, as the opening of the energy gap adds to the nonequilibrium excitations. 

Even with the advantage gained from the energy gap, the resulting irreversible dynamics on short timescales set a limit on the performance of the engine cycle. In fact, this problem does not just appear in dynamics about a critical point, but is present in any non-adiabatic driving of quantum heat engines. To improve engine performance on finite timescales different shortcut to adiabaticity (STA) approaches have been suggested \cite{GooldSciRep,Adolfo2012,delCampoEntropy,jing2,Abah2018,AbahPRE:19,Hartmann2020}. However, while STAs have been successfully developed for non-interacting and mean-field systems, designing them for strongly interacting many-body systems poses new challenges when scale invariance can not be exploited \cite{LewisSTA}, and is especially difficult due to the orthogonality catastrophe in larger systems \cite{polkovnikov,TF_PRL2020}. We therefore employ a variational approach which can find the optimal driving amplitude $V^\text{STA}_n(t)$ for each of the single particle functions $\psi_n(x,t)$ which are used in the Slater determinant to construct the many-body state \cite{Xu:20}. However, while this in principle can optimise the dynamics of each $\psi_n(x,t)$ individually, in practice a single lattice ramp must act on the entire many-body state and the chosen $V^\text{STA}_n(t)$ may  create unwanted excitations in different $\psi_m(x,t)$, for $m\neq n$.

We therefore consider two different approximate STAs to optimize the many-body dynamics. First, we choose the average of the STA pulses for all states up to the energy gap, $\langle V^\text{STA}(t) \rangle=\sum_{n=1}^M V^\text{STA}_n(t)/M$, and numerically time evolve the single particle states with this finite time ramp. While this shows an improvement over the non-optimized ramp $V_{\lambda}(t)$ for all timescales (see Fig.~\ref{fig:ref_ramp}(d-e)), it is only marginal as the optimization is averaged over the whole system. We therefore also consider the ramp $V^\text{STA}_{M-1}(t)$, which specifically optimizes the most irreversible single particle state, $\psi_{M-1}(x)$, which sits just below the gap and possesses the most excess energy after the $V_{\lambda}$ ramp (see insets in Fig.~\ref{fig:ref_ramp}(d-e)). This STA results in a larger efficiency gain as excitations of this state are mostly suppressed, and the adiabatic limit is quickly reached when the lattice is weak. However, this STA becomes ineffective for fast cycles, as large modulations in the approximate STA ramp can induce excitations in the rest of the system, which is a limitation of using these approximate techniques to design STAs for many-body states.

\section{Conclusions}

In summary, we have described the operation of a quantum Otto cycle about a critical point in a strongly interacting many-particle system. We have shown that such a setup can yield increased performance due to the presence of an energy gap and cooperative many-body effects which arise due to competition between interactions and lattice forces. Using the particular cold-atom setup we have chosen, which has already been experimentally studied \cite{haller2010pinning}, clearly highlights the dynamical effects stemming from the ordering when going through the critical point and the complex dynamics that arises during non-trivial shortcut driving. Furthermore, recent experiments have shown that many-particle heat engines can be realized with two-component ultracold gases \cite{Bouton2020}, whereby inelastic spin-exchange collisions are used to transfer heat between the engine and the bath. Accordingly, our work lays foundations for the further exploration of STA techniques for interacting many-body systems and their potential applications in quantum heat engines. 

\ack
%\section{Acknowledgements}
We thank Blaith\'in Power, John Goold and Steve Campbell for efficient discussions. This work was supported by the Okinawa Institute of Science and Technology Graduate University and JSPS KAKENHI-18K13507.
\vspace{0.3in}

\bibliography{TGEngine}
\section{Supplementary Material}
\subsection{Efficiency and power ratios}
In following we will describe how the efficiency and power ratios, $\eta_{AD}^*(N)$ and $P_{AD}^*(N)$, can be approximated in the deep lattice ($V_f\gg1$) and low temperature ($\theta=\frac{E_R}{k_B T_H}\sqrt{V_f} >1$) regime as given in Eq.~4 in the main text. 

For this we first look at the compression stroke for a system with $N$ particles at zero temperature ($T_C=0$). At the beginning of the compression stroke the particles fill a box of length $L$ which has single particle energies given by $\epsilon^B_n=(n+1)^2$. The groundstate energy of the $N$-particle system is therefore simply
\begin{equation}
    \langle H_{0}(0) \rangle=\sum_{n=0}^{N-1}\epsilon^B_n=(N+1)(2N+1)/6N.
\end{equation}
At the end of the compression stroke the lattice has a depth $V_f$ and at unit filling the lowest band is fully populated with one particle per lattice site. We can therefore treat this as $N$ isolated single particles confined to individual wells whose potential can be 
approximated by
\begin{equation}
    V \cos^2(k x+\pi/2)\approx V k^2 x^2-\frac{V}{3}k^4 x^4+\mathcal{O}(x^6)\;.
\label{eq:anharm}
\end{equation}
Treating the quartic term as a perturbation then gives the single particle energy spectrum in one deep lattice site as
\begin{equation}
    \epsilon^l_n\approx \left(n+\frac{1}{2}\right)2\sqrt{V_f}+\frac{1}{4}\left(2(n+1)-2(n+1)^2-1 \right),
    \label{eq:Epert}
\end{equation}
where $n=\{0,1,2,\dots\}$. Here the first term is the harmonic oscillator energy, $\hbar\omega=2\sqrt{V E_R}$ with the rescaled $V_f=V E_R$, and the second term is the correction due to the anharmonicity of the lattice site. The energy gap is the difference between two lowest single particle states, $\Delta=\epsilon^l_1-\epsilon^l_0=2\sqrt{V_f}-1$ and the total energy of $N$ particles in the lowest energy band is 
\begin{equation}
    \langle H_{0}(V_f) \rangle =N \epsilon^l_0=N\left(\sqrt{V_f}-\frac{1}{4}\right). 
    \label{EL0}
\end{equation}

Next we calculate the energy of the many-body state during the expansion stroke at temperature $T_H$. At the start of this stroke the lattice depth is fixed at $V_f$ and the partition function is well approximated by that of the harmonic oscillator
\begin{equation}
    \mathcal{Z}=\sum_{n=0}^{\infty}e^{-(2n+1)\theta}=\frac{\text{csch}(\theta)}{2}\;,
\end{equation}
which is justified when $\sqrt{V_f}\gg1/4$. %and it significantly eases the calculations of the efficiency and power of the Otto cycle shown in the following 
Alternatively, one may use the partition function as calculated from the energies given in Eq.~\eqref{eq:Epert}, resulting in
\begin{equation}
    \tilde{\mathcal{Z}}=\frac{\text{csch}(\theta)}{2}+\frac{\theta \coth^2(\theta)\text{csch}(\theta)}{8\sqrt{V_f}},
\end{equation}
which will yield qualitatively similar results to the low temperature harmonic oscillator approximation, however leads to more complex expressions. The total energy of the $N$ particle system is therefore $N$-times the single particle energy, $\langle H_{T_H}(V_f) \rangle=\frac{N}{\mathcal{Z}} \sum_{m=0}^{\infty} \epsilon^l_m e^{-(2m+1)\theta}$. This gives
\begin{equation}
    \langle H_{T_H}(V_f) \rangle=%\frac{M}{\mathcal{Z}} \sum_{n=0}^{\infty} \epsilon^L_n e^{-(2n+1)\theta}
    N\left( \sqrt{V_f}\coth(\theta)-\frac{1}{4}\coth^2(\theta)\right),
\end{equation}
where the first term describes the thermal state of a harmonic oscillator and the second term is the correction due to the anharmonicity of the lattice site.

%The thermal energy of $M$ particles without the lattice is also described by the band structure of the pinned system, however the single particle energies of each band are not degenerate but are described by $(n+1)^2$. We therefore must find the total energy of each $m^{th}$ band, $\tilde{\epsilon}^B_m$, and use this to calculate the $M$-body energy such that $E^B(\theta)=\frac{1}{\mathcal{Z}} \sum_{m=0}^{\infty} \tilde{\epsilon}^B_m e^{-(2m+1)\theta}$.
%%%%%%%%%%%%%%%%%%%%%%

Finally, at the end of the expansion stroke the lattice is removed and we must describe the thermal state of $N$ particles in the box potential. As the thermal statistics of the particles are still described by the lattice band structure we need to evaluate the total energy as a distribution over the each band. Since each band can contain $N$ particles which have single particle energies described by $(n+1)^2$, the total energy of the $m^{th}$ band is given by
\begin{equation}
    \tilde{\epsilon}^B_m=\frac{1}{6 N}\left[1+3 N\left(2m+1\right)+2N^2 \left(3m^2+3m+1 \right)\right]\;.
\end{equation}
Therefore, the total energy for $N$ particles is given by $\langle H_{T_H}(0) \rangle=\frac{1}{\mathcal{Z}} \sum_{m=0}^{\infty} \tilde{\epsilon}^B_m e^{-(2m+1)\theta}$ which gives
\begin{equation}
    \langle H_{T_H}(0)\rangle=\frac{1}{6N}\left(1+2N^2+3N\coth(\theta)+3N^2\text{csch}^2(\theta) \right)\;.
\end{equation}
The average work done during the compression and expansion strokes can then be calculated from $\langle W_C \rangle=\langle H_{0}(V_f)\rangle-\langle H_{0}(0)\rangle$ and $\langle W_H \rangle=\langle H_{T_H}(0)\rangle-\langle H_{T_H}(V_f)\rangle$ respectively, while the heat exchange with the hot bath is $\langle Q_H \rangle=\langle H_{T_H}(V_f) \rangle - \langle H_{0}(V_f) \rangle$. With these quantities the efficiency and power of the adiabatic mQHE can be calculated as in the main text.

Similarly the energies for the sQHE cycle can be straightforwardly calculated, with the energy of a single particle in the lattice potential simply given by $\langle H_{0}^{1}(V_f)\rangle =\langle H_{0}(V_f)\rangle/N$ and $\langle H_{T_H}^1(V_f)\rangle=\langle H_{T_H}(V_f)\rangle/N$, while in the box potential they are
\begin{align}
    \langle H_{0}^{1}(0)\rangle&=1\;,\\
    \langle H_{T_H}^{1}(0)\rangle&=\frac{1}{2}\coth(\theta)\left[1+\coth(\theta)\right]\;.
\end{align}
As above, the efficiency and power of the sQHE can then be calculated.

Finally, to compare the performance of the mQHE to the sQHE we calculate the ratios of the efficiency $\eta^*=\eta(N)/\eta(1)$ and power $P^*=P(N)/(N P(1))$. Using the above approximations for an engine operating in the deep lattice and low temperature regime we
find that we can write both ratios as
\begin{equation}
    \eta^*_{AD}(N)=P^*_{AD}(N)=1+\frac{1-1/N}{\Delta-\frac{3}{2}\left[ \coth\left(\theta\right)+1\right]}\;.
\label{eq:supp2}
\end{equation}

%Furthermore, increasing the number of particles and reaching the state of double filling, $N=2M$, where the two lowest states of each lattice site are occupied, does not lead to improved performance. In the limits $\theta,M\rightarrow\infty$ the efficiency ratio can be written as $\eta^*(2M)\equiv\frac{\eta(2M)}{\eta(1)}\rightarrow\frac{1-4(\Delta-1)^{-1}}{1-3\Delta^{-1}}$, showing that for $V_f>1$ the efficiency of the sQHE is always larger than the mQHE at double filling, which is due to the anharmonicity of the individual lattice sites leading to reduced gaps between higher lying energy states.  

\subsection{Shortcut to adiabaticity} 
 
To design an STA for the time-dependent ramp acting on the whole system we use a variational approach \cite{jing2,LewisSTA,Xu:20}. This method relies on minimizing the Lagrangian and finding the optimal $V_n(t)$ ramp for each single particle state individually. The success of this approach depends strongly on the choice of the ansatz for the time evolution of the corresponding single particle state. For our work we choose a simple ansatz of the form of a superposition between the the initial and the target state \cite{LewisSTA}
\begin{align}
    \Phi_n^c (x,t) &= \Psi_n(x,t) e^{i b(t) x^2}\\
    &= \mathcal{N}(t)\left[ (1-\varepsilon(t))\psi_n^I(x)+\varepsilon(t) \psi_n^F(x)\right] e^{i b(t) x^2}
\end{align}
where $\psi_n^I(x,t)$ is the initial $n^{th}$ single particle state and $\psi_n^F(x,t)$ is the corresponding target state, with $b(t)$ being a dynamical phase. We use the time dependent parameter $\varepsilon(t)$ to  switch the single particle state from its initial to the target state. To be able to obey the boundary conditions $\varepsilon(0)=0$, $\varepsilon(t_f)=1$, and  $\dot{\varepsilon}(0)=\dot{\varepsilon}(t_f)=\ddot{\varepsilon}(0)=\ddot{\varepsilon}(t_f)=0$, we choose a functional form of the switching parameter of $\varepsilon(t)=\sum_{j=0}^5 a_j t^j$. 

The resulting optimised lattice ramp is then given by
\begin{equation}
    V^\text{STA}_n(t)=-\frac{\frac{\partial \xi^2}{\partial \varepsilon}\left(\frac{\partial b}{\partial t}+2 b^2\right)+\frac{1}{2}\frac{\partial \beta}{\partial \varepsilon}}{\frac{\partial \alpha}{\partial \varepsilon}}\;,
\end{equation}
where the variables are given by 
\begin{align}
    \xi^2&=\int_{-L/2}^{L/2} x^2 |\Psi_n(x,t)|^2 dx\;,\\
    \alpha&=\int_{-L/2}^{L/2}\cos^2(k x) |\Psi_n(x,t)|^2 dx\;,\\
    \beta&=\int_{-L/2}^{L/2} \left|\frac{\partial \Psi_n(x,t)}{\partial x}\right|^2 dx\;,\\
    b&=\frac{1}{4 \xi^2}\frac{\partial \xi^2}{\partial t}\;.
\end{align}
Here $\xi^2$ is the width of the single particle state, $\alpha$ is the contribution from the lattice, $\beta$ the kinetic energy and $b$ is the chirp.

\end{document}